\documentclass[12pt]{article}
\usepackage{amssymb}
\usepackage{authblk}
\usepackage{amsmath}
\usepackage{hyperref}
\usepackage{graphicx}
\usepackage{subfig}
\usepackage{bm}
\usepackage{latexsym}
\usepackage{mathrsfs}
\usepackage{gensymb}
\begin{document}
\title{\bf Electromagnetic Duality Sensitivity of Holographic Complexity}
\author{Mojtaba Shahbazi\thanks{Corresponding author: mojtaba.shahbazi@modares.ac.ir}}
\author{Mehdi Sadeghi\thanks{mehdi.sadeghi@abru.ac.ir}}
\affil{Department of Physics, Faculty of Basic Sciences, Ayatollah Boroujerdi University, Boroujerd, Iran}
\date{\today}
\maketitle

\begin{abstract}
We investigate electromagnetic duality as a diagnostic of the sensitivity of information content of holographic complexity via "complexity=anything" in Einstein-ModMax theory. Functionals constructed solely from gravitational invariants are duality invariant, whereas matter-sensitive functionals can distinguish electromagnetic configurations that share the same bulk geometry and energy-momentum tensor. As an explicit example, we consider a functional involving $F_{\mu\nu}F^{\mu\nu}$ and show that its complexity and complexity growth vary along the duality orbit, interpolating between purely electric, mixed electric-magnetic, and purely magnetic configurations. This provides an explicit realization of the freedom inherent in generalized holographic complexity and shows that different choices of complexity functional can retain different sensitivity to information about the bulk matter sector even when the gravitational geometry is insensitive to that information.
\end{abstract}
\section{Introduction} \label{intro}
Electromagnetic duality is a fundamental symmetry of Maxwell theory that rotates the electric and magnetic sectors into each other. ModMax is a nonlinear extension of Maxwell theory that retains both conformal invariance and electromagnetic duality in four dimensions \cite{Bandos:2020jsw,ModMaxII}. The simultaneous preservation of conformal invariance and electromagnetic duality makes ModMax qualitatively different from generic nonlinear electrodynamics. The structure and symmetry properties of the theory have been further studied in \cite{Bandos2021a,Bandos2021b}, including its supersymmetric extension and its relation to generalized
Born-Infeld-type theories, in \cite{modsol} explicitly couples ModMax to Einstein gravity and constructs dyonically charged solutions and \cite{modsol1} studies black holes in Einstein-ModMax and magnetic effects.

ModMax has also received attention in holographic settings.
A dimensional reduction of four-dimensional Einstein-ModMax theory and construction of an
$AdS_2$/JT-gravity description including ModMax corrections has been considered in \cite{RathiRoychowdhury2023}. More recently, it has been studied in Einstein-ModMax holography in the
presence of momentum relaxation and an external magnetic field,
computing the DC transport coefficients and showing that the ModMax parameter modifies the longitudinal and Hall conductivities as well as the Nernst response \cite{Barrientos2025}. These results demonstrate that ModMax provides a useful holographic realization of nonlinear electromagnetic dynamics while retaining its characteristic
electromagnetic duality. 

A key question is whether physical observables can distinguish electromagnetic configurations related by electromagnetic duality. This question is particularly interesting when the electromagnetic field is coupled to gravity, since different electromagnetic configurations can produce the same gravitational geometry. When coupled to gravity, it admits charged AdS black brane solutions whose electromagnetic configurations can be continuously related by duality transformations \cite{Kruglov,KRUGLOV2015299,Kar:2024zbo}. In particular, a purely electric configuration can be transformed into a dyonic or purely magnetic one without changing the corresponding gravitational solution.

Holographic complexity provides a useful probe of this distinction. The standard CV, CA and CV2.0 proposals associate complexity with different geometric quantities in the bulk \cite{cv,ca,cv2}. Holographic
complexity has been investigated for Born-Infeld black holes using both the CA and CV proposals \cite{TaoWangYang2017,Meng2019}, while subregion complexity has also been studied in Einstein-Born-Infeld backgrounds \cite{PanJing2019,ShiPanJing2020}. More generally, the effects of nonlinear electromagnetic interactions on holographic complexity have been investigated in holographic superconductors with Born-Infeld, logarithmic, and exponential nonlinear electrodynamics \cite{LingLiuWu2021}. These studies demonstrate that nonlinear electromagnetic interactions can leave nontrivial signatures in holographic complexity and establish complexity as a potentially useful probe of the bulk matter sector.

The "complexity equals anything'' (CAny) proposal allows a broad class of codimension-one functionals \cite{general},
\begin{equation}
\mathcal O_{F_1,\Sigma_{F_2}}=
\frac{1}{G_N L}
\int_{\Sigma_{F_2}}d^d\sigma\sqrt{h}
F_1(g_{\mu\nu};X^\mu),
\end{equation}
This functional freedom raises a natural question:can holographic complexity distinguish electromagnetic configurations with identical gravitational geometry?

The answer depends on the choice of $F_1$. If $F_1$ is constructed only from curvature invariants, its value is unchanged when the metric and the extremized hypersurface are unchanged. Such complexity functionals are therefore invariant along the electromagnetic duality orbit. In contrast, a matter-sensitive choice such as
\begin{equation}
F_1=F_{\mu\nu}F^{\mu\nu}
\end{equation}
can depend explicitly on the electromagnetic configuration and hence distinguish duality-related solutions.

We quantify this distinction by defining the orbit complexity
\begin{equation}
\Delta_{R} \mathcal C_{Any}=\mathcal C_{Any}^{R(\theta)}-\mathcal C_{Any}^{R(0)},
\end{equation}
which measures the response of complexity to an electromagnetic duality rotation. For the Einstein-ModMax black brane, we compare a curvature-dependent functional involving the Weyl tensor squared with the matter-sensitive electromagnetic functional. While the former remains invariant along the duality orbit, the latter depends nontrivially on the electromagnetic configuration. We further show that the corresponding effective potential for the complexity growth can change its structure under duality rotations, leading to different branches of the complexity growth rate where there are critical duality angles $\theta_c$, in the theory that the change in the number of branches in the complexity growth happen.

From the boundary perspective, the duality-related solutions have the same bulk metric and therefore the same boundary metric and the bulk energy-momentum tensor. Nevertheless, their asymptotic electromagnetic data are different: a duality rotation generates magnetic data from the original electric configuration. Thus, matter-sensitive complexity can distinguish electromagnetic configurations that are degenerate with respect to the gravitational geometry and stress tensor.

These results show that the functional freedom of holographic complexity has a direct physical interpretation. Curvature-based functionals probe the gravitational sector and are insensitive to the electromagnetic duality orbit, whereas matter-sensitive functionals are sensitive to information about the underlying electromagnetic configuration.

The paper is organized as follows. In the sec.~\ref{sec2}, we introduce Einstein-ModMax theory and construct the electric charged AdS black brane solution. In the sec.~\ref{sec3}, we calculate the generalized holographic complexity and its growth rate. In the sec.~\ref{sec4}, we study electromagnetic duality and its effect on the complexity functionals and their effective potential. The section \ref{numer} provides the numerically analysis of the complexity of the duality rotated solution and introduces critical duality angles that change the structure of the complexity growth rate branches. In the section \ref{bound} the boundary perspective of the duality rotation is discussed. Finally, the sec.~\ref{con} summarizes our results and discusses their consequences.

\section{Einstein-ModMax theory}\label{sec2}
The four-dimensional action for the modification of Maxwell theory in an Anti-de Sitter (AdS) background, incorporating a negative cosmological constant, is expressed as \cite{Kruglov, KRUGLOV2015299, Kar:2024zbo}
\begin{eqnarray}\label{action}
	S=\frac{1}{16\pi G}\int d^{4}x \sqrt{-g} \bigg[ R - 2\Lambda + \mathcal{L} \bigg],
\end{eqnarray}
where $R$ is the Ricci scalar, and $\Lambda$ is related to the AdS radius $l$ by $\Lambda = -\frac{3}{l^2}$.

In addition, the Lagrangian density $\mathcal{L}$ corresponds to the modified Maxwell Lagrangian \cite{Bandos:2020jsw, ModMaxII}, assumed to be comparable in three and four-dimensional spacetimes
\begin{equation} \label{ModMaxL}
	\mathcal{L} = -\mathcal{F} \cosh \gamma + \sqrt{\mathcal{F}^2 + \widetilde{\mathcal{F}}^2} \sinh \gamma,
\end{equation}
where $\gamma$ is a dimensionless parameter termed the ModMax parameter. 

In this formulation, $\mathcal{F} =\frac{1}{4} F_{\mu \nu} F^{\mu \nu}$ is the scalar invariant, and $\widetilde{\mathcal{F}} =\frac{1}{4} F_{\mu\nu} \tilde{F}^{\mu\nu}$ is the pseudoscalar, with
\begin{equation}
	\tilde{F}^{\mu \nu} = \frac{1}{2} \epsilon^{\mu\nu\rho\lambda} F_{\rho \lambda}.
\end{equation}
Notably, when $\gamma = 0$, the ModMax Lagrangian reduces to the conventional linear Maxwell form: $\mathcal{L} = -\mathcal{F}$ \cite{Bandos:2020jsw,ModMaxII}.

A static metric ansatz of the form
\begin{equation}\label{metric1}
	ds^{2} = -f(r) dt^{2} + \frac{dr^{2}}{f(r)} + \frac{r^{2}}{l^{2}} (dx^{2} + dy^{2}),
\end{equation}
is adopted, where $f(r)$ is an unknown function of the radial coordinate $r$, and $l$ is related to the cosmological constant.

The gauge field $A_\mu$ is expressed as
\begin{equation}\label{background}
	A_\mu=h(r) dt
\end{equation}
Varying the action, Eq.~(\ref{action}), with respect to the metric \(g^{\mu\nu}\) and the gauge field \(A^{\nu}\), respectively, yields the following field equations:
\begin{equation}
	R_{\mu\nu} - \frac{1}{2} g_{\mu\nu} R + \Lambda g_{\mu\nu}
	- 2 e^{-\gamma} F_{\mu\alpha} F_{\nu}^{\alpha}
	+ \frac{1}{2} e^{-\gamma} F_{\alpha\beta} F^{\alpha\beta} g_{\mu\nu} = 0,
\end{equation}
and
\begin{equation}\label{Max}
	\nabla_\mu \left( e^{-\gamma} F^{\mu\nu} \right) = 0.
\end{equation}
Substituting Eq.~(\ref{metric1}) into Eq.~(\ref{Max}) gives,
\begin{equation}
	2 h'(r) + r h''(r) = 0.
\end{equation}
The general solution for \(h(r)\) is,
\begin{equation}\label{h}
	h(r) = C_2 - \frac{C_1}{r},
\end{equation}
where \(C_1\) and \(C_2\) are integration constants.\\
 Applying the regularity condition at the event horizon, \(A_t(r_h) = 0\), yields \(C_2 = C_1/r_h\). Finally,
\begin{equation}
	h(r) = Q \left( \frac{1}{r} - \frac{1}{r_h} \right),
\end{equation}
where \(Q\) characterizes the charge associated with the gauge field.\\
The $tt$-component of the Einstein equation is as,
\begin{equation}
	f(r) + \Lambda r^2 + r f'(r) + e^{-\gamma}  r^2 h'(r)^2=0
\end{equation}
so $f(r)$ is as,
\begin{equation}\label{sol}
	f(r) = - \frac{2m}{r} - \frac{\Lambda r^2}{3} +  \frac{Q^2 }{r^2}e^{-\gamma},
	\end{equation}
The invariant scalar $\mathcal{F}$ is obtained via
\begin{equation}
	F_{tr} = - F_{rt} = \frac{Q}{r^{2}}.
\end{equation}
The event horizon, located at $r = r_h$, satisfies $f(r_h) = 0$, fixing $m$ as
\begin{equation}
	m = \frac{r_h^{3}}{2 l^{2}} + \frac{Q^2}{2 r_h}  e^{-\gamma}.
\end{equation}
Thus, the metric function becomes:
\begin{equation}\label{black}
	f(r) = \frac{r^2}{l^2} \left( 1 - \frac{r_h^{3}}{r^{3}} \right) + \frac{Q^2}{r^2}  \left( 1 - \frac{r}{r_h} \right) e^{-\gamma}.
\end{equation}
The Hawking temperature, associated with the surface gravity at the horizon, is given by
\begin{align}\label{Temp}
	T = \frac{f'(r_h)}{4 \pi} = \frac{r_h}{4 \pi l^2} + \frac{Q^2}{4 \pi r_h^3}e^{-\gamma}.
\end{align}

\section{$\mathcal C_{Any}$ on the black brane solution}\label{sec3}
Complexity equals to anything conjecture has emerged as a generalized version of complexity in bulk. In this proposal complexity is not limited to a specific geometric quantity but can be described by a more general function to include a class of new diffeomorphism invariant observables \cite{general}. Here we concentrate on the codimension-one of bulk region cases
\begin{align}\label{obser}
	\mathcal{O}_{F_1,\Sigma_{F_2}}(\Sigma_{CFT})=\frac{1}{G_N L}\int_{\Sigma_{F_2}}\mathrm{d}^{d}\sigma \sqrt{h} F_1(g_{\mu\nu};X^{\mu})
\end{align}
where, $F_1$ can equal to one which led to the volume of hypersurface as in CV. It can be general scalar function of metric $g_{\mu\nu}$  and an embedding $X^{\mu}(\sigma^a)$ of the hypersurfaces. Furthermore, $\Sigma_{F_2}$ is codimension-one hypersurface in the bulk spacetime with boundary time slice $\partial\Sigma_{F_2} =\Sigma_{CFT}$. Extremality of the hypersurface gives
\begin{align}\label{var}
	\delta_{X}\Big(\int_{\Sigma}\mathrm{d}^{d}\sigma \sqrt{h} F_2(g_{\mu\nu};X^{\mu})\Big)=0. 
\end{align}
For simplicity, we follow the case $F_1 =F_2$, so the observable \eqref{obser}, known by $\mathcal{C}_{Any}$, obeying above condition is expressed by
\begin{align}\label{maxv}
	\mathcal{C}_{Any}(\tau)= \max_{\partial\Sigma(\tau)=\Sigma_{\tau}}\frac{V_x}{G_N L}\left[\int_{\Sigma}\mathrm{d}^{d}\sigma \sqrt{h} F_1(g_{\mu\nu};X^{\mu})\right]
\end{align}
where $h$ is the determinant of induced metric on the given hypersurface.
The usual choice of generalization is
\begin{align}\label{F1F2}
	F_1=F_2=1+\alpha L^4 \mathcal{C}^2,
\end{align}
where $\mathcal{C}^2$ is the Weyl tensor squared and for $\alpha=0$ the CV is retrieved. In the following, we are going to study the $\mathcal{C}_{Any}$ of this setup, with the solution \eqref{metric1}. With this aim, we need to transform the metric to Eddington-Finkelstein coordinates
\begin{align}\label{metricEF}
	\mathrm{d} s^{2}=-f(r) \mathrm{d} v^{2}+ 2\mathrm{d} v \mathrm{d} r+ r^2 d\mathcal{S}^2,
\end{align}
where
\begin{align}
	v=t+r^{*}(r), \quad \mathrm{d}r^{*}=\frac{\mathrm{d}r}{f(r)},	
\end{align}
The generalized volume complexity as a codimension-one observable could be obtained from \eqref{maxv} and \eqref{F1F2} as \cite{general}
\begin{align}\label{Any}
	\mathcal{C}_{Any}(\tau)= \frac{V_0}{G_N L}\int_{\Sigma}\mathrm{d}\sigma  r^{d-1}\sqrt{-f \dot{v}^{2}+2 \dot{v} \dot{r}} a(r)
\end{align}
Here, the dots indicate derivatives with respect to $\sigma$, and $V_0$ represents the volume of spatial directions $x$ and $y$.

Considering $\mathcal{C}_{Any}$ as an action, where extremizing it is equivalent to solving its equations of motion, leads to a conserved  momentum conjugate to coordinate $v$, because of  spacetime is stationary 
\begin{align}
	P_v=-\frac{\partial \mathcal{L}}{\partial \dot{v}}=\frac{a(r)r^{d-1}(\dot{r}-f(r) \dot{v})}{\sqrt{-f \dot{v}^{2}+2\dot{v} \dot{r}}}=\dot{r}-f(r) \dot{v}.	
\end{align}
Since \eqref{Any} is diffeomorphism invariant and doesn't change under reparametrization, one can fix parameter $\sigma$ by choosing 
\begin{align}
	\sqrt{-f \dot{v}^{2}+2\dot{v} \dot{r}}=a(r)r^{d-1}.
\end{align}
Then it is straightforward to derive extremality conditions
\begin{align}\label{rdot}
	\dot{r}=\pm \sqrt{P^2_v +f(r)a(r)^2r^{2(d-1)}},
\end{align}
\begin{align}\label{tau}
	\dot{t}=\dot{v}-\frac{\dot{r}}{f(r)}=\frac{-P_v\dot{r}}{f(r) \sqrt{	P^2_v- \tilde{U}(r)}}.
\end{align}
where $\tilde{U}(r)$ taking effective potential
\begin{align}\label{upot}
	\tilde{U}(r)=-f(r)a(r)^2r^{2(d-1)}.
\end{align}
then, \eqref{rdot} form equation of motion like for a classical particle	
\begin{align}\label{eom}
	\dot{r}^{2}+\tilde{U}(r)=P^2_v.
\end{align}
 In the symmetric trajectory, the conserved momentum is a function of turning point $r_{min}$
	\begin{align}
		P^2_v=	\tilde{U}(r_{min})=-f(r_{min})a(r_{min})^2r_{min}^{2(d-1)}.
	\end{align}	
	On the other hand, by integrating \eqref{tau}, the boundary time relate to conserved momentum and turning point $r_{min}$ act as an intermediary
	\begin{align}\label{tauint}
		\tau & =2 \int_{r_{\min }}^{\infty} \mathrm{d} r \frac{-P_v}{f(r) \sqrt{	P^2_v -\tilde{U}(r)}}.
	\end{align}
	 Boundary time derivation of $\mathcal{C}_{Any}$ leads to 
	\begin{align}
		\frac{\mathrm{d} \mathcal{C}_{Any}}{\mathrm{~d} \tau}=\frac{1}{2} \frac{\mathrm{d} \mathcal{C}_{Any}}{\mathrm{~d} \tau_{\mathrm{R}}}&=\frac{V_0}{G_N L} P_v=\frac{V_0}{G_NL} \sqrt{-f(r_{min})}a(r_{min})r_{min}^{(d-1)}.
	\end{align}
	So, growth rate of generalized complexity can be studied by the behavior of conserved momentum in full-time. Especially, the late-time behavior of growth rate is determined by 
	\begin{align}\label{tauinf}
		\lim _{\tau \rightarrow \infty} \frac{\mathrm{d} \mathcal{C}_{Any}}{\mathrm{~d} \tau}=\frac{V_0}{G_NL} \sqrt{-f\left(\tilde{r}_{\min }\right)} a(\tilde{r}_{\min })\tilde{r}_{min}^{(d-1)},
	\end{align}
	where $\tilde{r}_{\min }$ is local maximum of the effective potential, and because at this radius time goes to infinity, it is known sometimes as $r_f\equiv \tilde{r}_{\min }$.

\section{Electromagnetic-duality classification of $\mathcal C_{Any}$}\label{sec4}
ModMax theory is invariant under the electromagnetic duality \cite{Bandos:2020jsw}
\begin{equation}\label{mat}
\begin{pmatrix}
G'_{\mu\nu}\\[2mm]
\tilde F'_{\mu\nu}
\end{pmatrix}
=
\begin{pmatrix}
\cos\theta & \sin\theta\\
-\sin\theta & \cos\theta
\end{pmatrix}
\begin{pmatrix}
G_{\mu\nu}\\[2mm]
\tilde F_{\mu\nu}
\end{pmatrix}.
\end{equation}
where 
\begin{align}
G_{\mu\nu}=&-2\frac{\partial \mathcal L}{\partial F^{\mu\nu}}\\
=&-\Big(\cosh \gamma +\frac{\mathcal F \sinh \gamma}{\sqrt{\mathcal F^2+\widetilde{\mathcal F}^2}}\Big) F_{\mu\nu}-\frac{\widetilde{\mathcal F}\sinh \gamma}{\sqrt{\mathcal F^2+\widetilde{\mathcal F}^2}}\tilde F_{\mu\nu}
\end{align}
$\tilde F_{\mu\nu}$ is hodge dual of $F_{\mu\nu}$. We denote \eqref{mat} as $[F']=R(\theta)[F]$. The gravitational bulk solutions under the $R$ transformation remain intact. These transformations can classify the complexity functional $F_1$ into two categories:
\begin{itemize}
\item $R$-invariant: $F_1$ which is a function of $R$-invariant quantities $\rightarrow \mathcal C_{Any}(F_1)=\mathcal C_{Any}(F'_1)$\\
\item $R$-variant: $F_1$ which is a function of $R$-variant quantities $\rightarrow \mathcal C_{Any}(F_1)\neq\mathcal C_{Any}(F'_1)$
\end{itemize}
Examples of the first category are geometric invariant quantities such as $C^2$, the Weyl squared tensor, $R_{\mu\nu}R^{\mu\nu}$, the Ricci tensor squared, non-geometric quantities such as $\mathcal F^2+\widetilde{\mathcal F}^2$, mixed quantities such as $C^2(\mathcal F^2+\widetilde{\mathcal F}^2)$ and the second category is $F_{\mu\nu}F^{\mu\nu}$.

This categorization shows that complexity can distinguish the duality that gravity is insensitive to the position along the electromagnetic duality orbit, because of the fact that under an $R$ transformation, the gravitational bulk solution: the bulk energy-momentum tensor $T_{\mu\nu}$ and the metric $g_{\mu\nu}$ are intact. The gravitational geometry is invariant along the electromagnetic duality orbit, but a matter-sensitive notion of complexity is not because of the fact the the embedding $X^{\mu}$ changes in an $R$-variant functional. This means that the freedom of the complexity functional $F_1$ as an arbitrary ambiguity could be interpreted as the different information content of complexity. In other words, some complexity functionals see the duality invariant content and some do not.

\subsection{Orbit complexity}
By the behavior of the complexity under the $R$ transformation we can define a difference in complexity under this transformation as follows
\begin{align}
\Delta_R \mathcal C_{Any}:= \mathcal C_{Any}^{R(\theta)}-\mathcal C_{Any}^{R(0)}.
\end{align}
We call this quantity \textit{orbit complexity}. The orbit complexity of the first category ($R$-invariant) vanishes, while in the second category it is nonvanishing. This behavior shows that the duality response of the complexity can represent the sensitivity of the complexity functional to the duality as follows
\begin{align}
\mathcal S_{F_1}:=\frac{\partial \mathcal C_{Any}(F_1)}{\partial \theta}.
\end{align}
$\mathcal S$ could measure the sensitivity of the chosen functional to the electromagnetic duality. The orbit complexity can be expanded as,
\begin{align}
\Delta_R \mathcal C_{Any}\sim \mathcal S_{F_1}\theta+\cdots .
\end{align}

Suppose that the matter content of a theory is fixed, then by the Einstein equation (and boundary conditions) and the energy-momentum tensor, the metric is specified. 
Now, by an $R$ transformation, the metric and energy-momentum tensor are fixed, but the electromagnetic matter content changes. Then, an $R$-variant complexity functional can distinguish these two cases, while an $R$-invariant is invariant along the duality orbit. This means that the $R$-variant complexity functional retains the duality that the geometry alone can not.

\section{Numerical analysis and critical structure}\label{numer}
For the black brane solution considered in this work, we take the Abelian gauge potential
\begin{equation}
A=A_t(r) dt,
\end{equation}
with
\begin{equation}
A_t(r)=Q\left(\frac{1}{r_h}-\frac{1}{r}\right).
\end{equation}
The gauge has been chosen such that
\begin{equation}
A_t(r_h)=0.
\end{equation}
The corresponding field strength is \footnote{We set $l=1$ in the metric.}
\begin{align}
&F_{tr}=-\partial_r A_t=\frac{Q}{r^2},\\
&\epsilon_{xytr}=\sqrt{-g}=r^2,\\
&\tilde F_{xy}=-Q,\\
&\tilde F^{xy}=-\frac{Q}{r^4},\\
&\tilde F^2=\tilde F_{\mu\nu}\tilde F^{\mu\nu}=-F^2
\end{align}
and
\begin{align}
F_{\mu\nu}F^{\mu\nu}=-\frac{2Q^2}{r^4}
\end{align}
Then 
\begin{align}\label{gf}
&\mathcal F=-\frac{Q^2}{2r^4},\\
&\widetilde{\mathcal F}=0,\\
&G_{\mu\nu}=-e^{-\gamma}F_{\mu\nu}.
\end{align}

The electromagnetic duality transformation can then be expressed as a $SO(2)$ rotation in the space of electromagnetic field strengths,
\begin{align}
G'_{\mu\nu}&=G_{\mu\nu}\cos\theta+\tilde F_{\mu\nu}\sin\theta,\\
\tilde F'_{\mu\nu}&=-G_{\mu\nu}\sin\theta+\tilde F_{\mu\nu}\cos\theta.\label{tilf}
\end{align}
By Hodge star of \eqref{tilf}
\begin{align}
F'_{\mu\nu}=F_{\mu\nu}\cos \theta-e^{-\gamma}\tilde F_{\mu\nu} \sin \theta,
\end{align}
consequently,
\begin{align}\label{f'}
F'^2&=g^{\mu\rho}g^{\nu\sigma}F'_{\mu\nu}F'_{\rho\sigma}=F^2\cos^2\theta+e^{-2\gamma}\tilde F^2\sin^2\theta+2e^{-\gamma}F_{\mu\nu}\tilde F^{\mu\nu}\cos\theta \sin\theta,\\
&=-\frac{2Q^2}{r^4}(\cos^2\theta -e^{-2\gamma}\sin^2\theta)
\end{align}
In \eqref{f'}, the result reduces to Maxwell when $\gamma \rightarrow 0$ where $F'^2=-\frac{2Q^2}{r^4}\cos2\theta$. Setting $\theta=0$, it gives $F'^2=-\frac{2Q^2}{r^4}$, the pure electric field and setting $\theta=\frac{\pi}{2}$, gives $F'^2=\frac{2Q^2}{r^4}$, the pure magnetic field.

\begin{figure}[!h]
\centering
\subfloat[]{\includegraphics[width=5cm]{{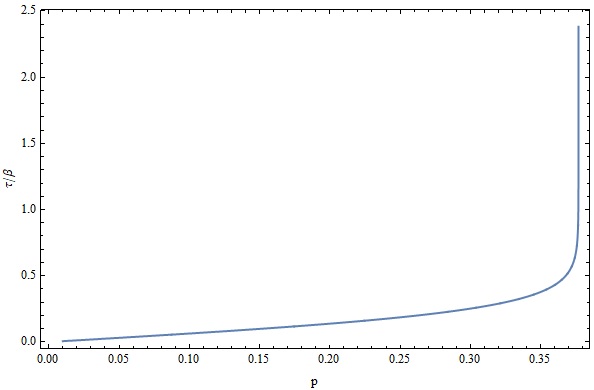}}\label{cgr1,0}}
\subfloat[]{\includegraphics[width=5cm]{{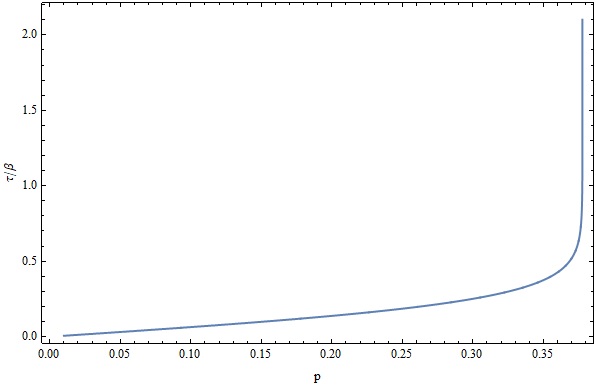}}\label{cgr1,p4}}
\subfloat[]{\includegraphics[width=5cm]{{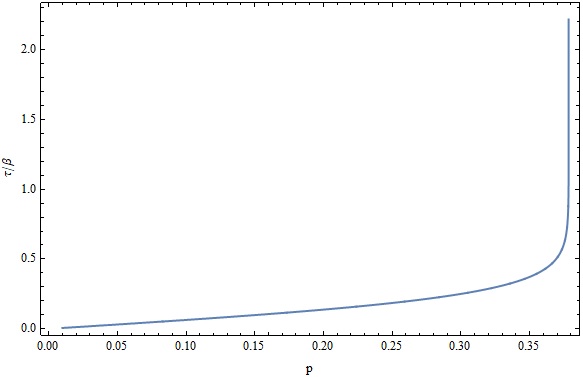}} \label{cgr1,p2}}\\
\subfloat[]{\includegraphics[width=5cm]{{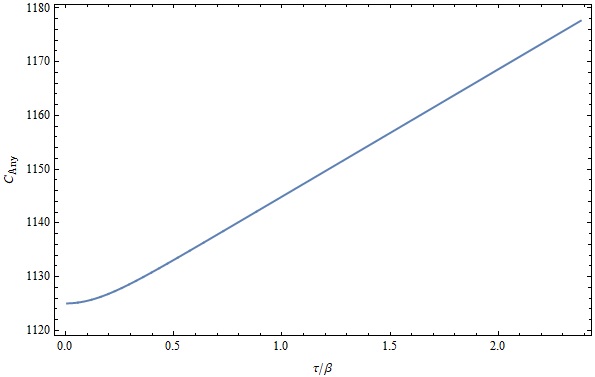}}\label{c1,0}}
\subfloat[]{\includegraphics[width=5cm]{{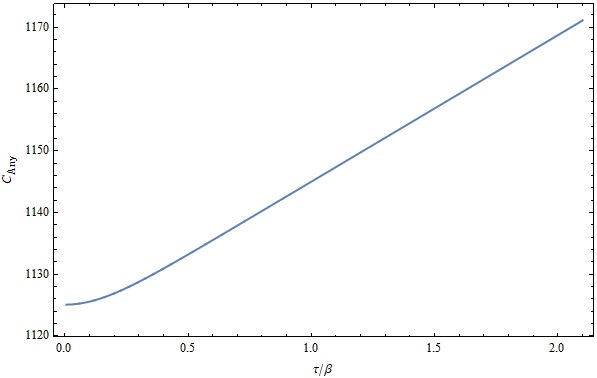}}\label{c1,p4}}
\subfloat[]{\includegraphics[width=5cm]{{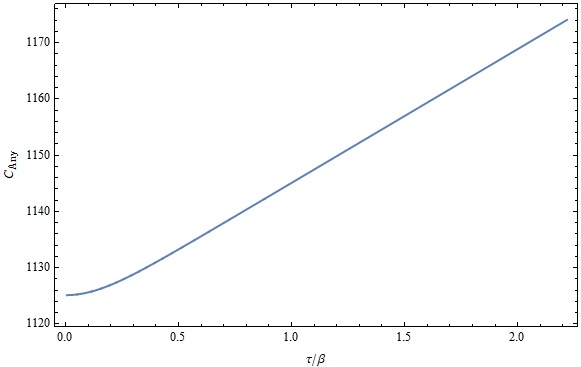}} \label{c1,p2}}
	\caption{The above panel depicts the complexity growth rate for a) $\theta=0$, b)$\theta=\frac{\pi}{4}$, c) $\theta=\frac{\pi}{2}$ and their complexities, respectively. The other parameters are set as $Q=1$, $\gamma=1$, the ModMax parameter and $\alpha=-0.002$, the generalized parameter. In the case $\theta=0$, there appears a pure electric field, in the case $\theta=\frac{\pi}{4}$, a mixed electric and magnetic field and $\theta=\frac{\pi}{2}$, a pure magnetic field.}\label{fig1}
\end{figure}

\begin{figure}[!h]
\centering
\includegraphics[width=10cm]{{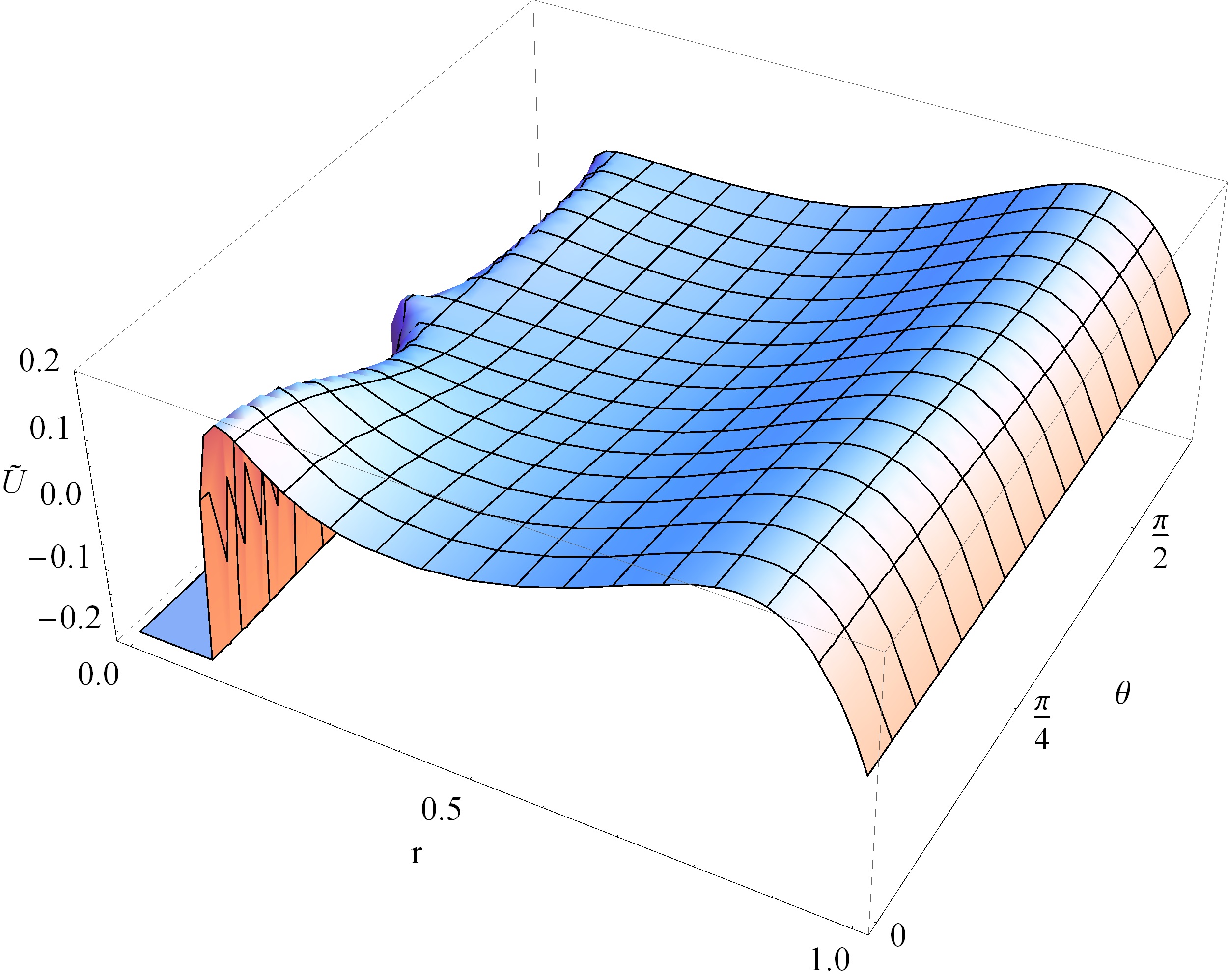}}
\caption{The effective $\tilde U$ potential in terms of $r$ and $\theta$. The other parameters are set $Q=0.5$, $\gamma=0.6$ and $\alpha=-0.1$. This diagram shows that the effective potential in the pure electrical content ($\theta=0$) has two maxima and one maximum near $\theta\approx 61^\circ$ and near $118^\circ$.}\label{fig2}
\end{figure}

In Fig. \ref{fig1}, the complexity and the complexity growth rate of functional $F_{\theta=0}^2$ and $F^2_{\theta}=F'^2$ are displaced. Although, Fig. \ref{fig1} shows, their qualitative behavior are the same (the number of branches) and their values differ, for different parameters it could be demonstrated that the number of branches changes. This fact could be examined by the shape of the effective $\tilde U$ potential in \eqref{upot} as depicted in Fig. \ref{fig2}. This means that there are some critical $\theta_c$ where the rotation makes the change in the number of branches. In our solution, the critical angles are near $\theta_c\approx 61^\circ$ and near $118^\circ$, where before the critical angles there are two maxima in the effective $\tilde U$ which leads to two branches in the complexity rate. Moreover, at the critical angle, there appears one maximum, leading one branch in the complexity rate. In the following section we derive a formula that finds the critical angle.

\subsection{Critical points of the effective potential}
To analyze the branch structure of the complexity rate the study of the maxima of the effective potential are crucial \cite{shahbazi, myers}. We consider the matter-sensitive choice \footnote{It is worth mentioning that the correct formula is $a=1+\alpha L F'^2$, but we set $L=1$.},
\begin{equation}
a(r,\theta)=1+\alpha F'^2(r,\theta),
\end{equation}
where the electromagnetic invariant along the duality orbit is
\begin{equation}
F'^2(r,\theta)=-\frac{2Q^2}{r^4}
\left(
\cos^2\theta-e^{-2\gamma}\sin^2\theta
\right).
\end{equation}
It is useful to introduce the parameter
\begin{equation}
B(\theta):=2\alpha Q^2
\left(
\cos^2\theta-e^{-2\gamma}\sin^2\theta
\right),
\end{equation}
such that
\begin{equation}
a(r,\theta)=1-\frac{B(\theta)}{r^4}.
\end{equation}

For the planar black brane geometry considered here, the effective potential governing the late-time extremization problem is
\begin{equation}
\tilde{U}(r,\theta)=-f(r)a^2(r,\theta)r^4.
\end{equation}
Consequently,
\begin{equation}
\tilde{U}(r,\theta)=-f(r)r^4
\left(
1-\frac{B(\theta)}{r^4}
\right)^2
=
-f(r)\frac{(r^4-B(\theta))^2}{r^4}.
\end{equation}
The stationary points of the effective potential are determined by
\begin{equation}
\frac{\partial\tilde{U}}{\partial r}=0.
\end{equation}
Using
\begin{equation}
a'(r,\theta)=\frac{4B(\theta)}{r^5},
\end{equation}
we obtain
\begin{align}
\frac{\partial\tilde{U}}{\partial r}
&=
-\Big(
f'r^4a^2
+4fr^3a^2
+2fr^4aa'
\Big)
\nonumber\\
&=
-ar^3
\Big(
a(rf'+4f)
+\frac{8B}{r^4}f
\Big).
\end{align}
Thus, away from the zeros of $a(r,\theta)$, the nontrivial stationary
points satisfy
\begin{equation}
\left(1-\frac{B}{r^4}\right)(rf'+4f)
+\frac{8B}{r^4}f
=0,
\end{equation}
or equivalently,
\begin{equation}
rf'+4f
+\frac{B}{r^4}(4f-rf')
=0.
\label{statg}
\end{equation}
For the black brane solution,
\begin{equation}
f(r)=
\frac{r^2}{l^2}
-\frac{r_h^3}{l^2r}
+\frac{Q^2e^{-\gamma}}{r^2}
-\frac{Q^2e^{-\gamma}}{r_hr},
\end{equation}
we define
\begin{equation}
K:= Q^2e^{-\gamma}.
\end{equation}
The derivative of the blackening function is then
\begin{equation}
f'(r)
=
\frac{2r}{l^2}
+\frac{r_h^3}{l^2r^2}
-\frac{2K}{r^3}
+\frac{K}{r_hr^2}.
\end{equation}
The combinations appearing in \eqref{statg} are
\begin{align}
rf'+4f
&=
\frac{6r^2}{l^2}
-\frac{3r_h^3}{l^2r}
+\frac{2K}{r^2}
-\frac{3K}{r_hr},
\\
4f-rf'
&=
\frac{2r^2}{l^2}
-\frac{5r_h^3}{l^2r}
+\frac{6K}{r^2}
-\frac{5K}{r_hr}.
\end{align}
Substitution into \eqref{statg} gives
\begin{equation}
\frac{6r^2}{l^2}
-\frac{3r_h^3}{l^2r}
+\frac{2K}{r^2}
-\frac{3K}{r_hr}
+
\frac{B}{r^4}
\left(
\frac{2r^2}{l^2}
-\frac{5r_h^3}{l^2r}
+\frac{6K}{r^2}
-\frac{5K}{r_hr}
\right)
=0.
\end{equation}
Multiplying by $l^2r^5r_h$, the stationary-point condition can be written
as the seventh-order polynomial equation
\begin{align}
P(r;B)
:=&
6r^7r_h
-3r^4r_h^4
+2Kl^2r^3r_h
-3Kl^2r^4
\nonumber\\
&+
B\left(
2r^3r_h
-5r^2r_h^4
+6Kl^2r_h
-5Kl^2r
\right)
=0.
\label{statp}
\end{align}
Since
\begin{equation}
B=B(\theta),
\end{equation}
the coefficients of this polynomial vary continuously along the
electromagnetic duality orbit. Therefore, the number and locations of its real roots can change as a function of the duality angle.

A change in the number of stationary points occurs when two stationary points merge. At such a critical point, the effective potential has a degenerate extremum. Consequently, the critical values $(r_c,B_c)$ satisfy
simultaneously
\begin{equation}
\frac{\partial\tilde{U}}{\partial r}
(r_c,B_c)
=0,
\qquad
\frac{\partial^2\tilde{U}}{\partial r^2}
(r_c,B_c)
=0.
\end{equation}
Equivalently, for the polynomial in \eqref{statp},
\begin{equation}
P(r_c;B_c)=0,
\qquad
\frac{\partial P}{\partial r}(r_c;B_c)=0.
\label{criticalc}
\end{equation}

These equations provide an analytic criterion for the transition between different numbers of branches.

Since \eqref{statp} is linear in $B$, it can be written
as
\begin{equation}
P(r;B)
=
P_0(r)+BP_1(r),
\end{equation}
where
\begin{align}
P_0(r)
={}&
6r^7r_h
-3r^4r_h^4
+2Kl^2r^3r_h
-3Kl^2r^4,
\\
P_1(r)
={}&
2r^3r_h
-5r^2r_h^4
+6Kl^2r_h
-5Kl^2r.
\end{align}

The first criticality condition gives
\begin{equation}
B_c
=
-\frac{P_0(r_c)}{P_1(r_c)},
\label{bcritical}
\end{equation}
provided $P_1(r_c)\neq0$. Substitution into the second condition yields an
equation involving only $r_c$,
\begin{equation}
P_0'(r_c)P_1(r_c)
-
P_0(r_c)P_1'(r_c)
=0.
\label{rcritical}
\end{equation}

Solving \eqref{rcritical} determines the radial position at which
two stationary points coalesce. The corresponding critical value $B_c$ then follows from \eqref{bcritical}. Finally, using
\begin{equation}
B(\theta)=2\alpha Q^2
\left(
\cos^2\theta-e^{-2\gamma}\sin^2\theta
\right),
\end{equation}
the corresponding critical duality angle is determined by
\begin{equation}
\cos^2\theta_c=\frac{B_c/(2\alpha Q^2)+e^{-2\gamma}}{1+e^{-2\gamma}}.
\end{equation}

Therefore, the transition in the number of branches can be characterized without relying solely on a numerical scan of the duality angle, it occurs when the polynomial $P(r;B(\theta))$ develops a multiple root, equivalently when \eqref{criticalc} are simultaneously satisfied. For parameters of Fig. \ref{fig2} ($Q=0.5$, $\gamma=0.6$ and $\alpha=-0.1$), critical angles are given
\begin{align}
B_c&=-2.09385497\times 10^{-6},\\
\cos^2\theta_c&=0.2315074,\\
\theta_c&=61.2393^\circ~and~118.7607^\circ
\end{align}

\section{Boundary interpretation}\label{bound}
The equality of the bulk metrics implies equality of the boundary metric source and, through holographic renormalization, equality of the stress tensor \footnote{We should note that the expectation value of the boundary energy $\langle T_{ij}\rangle$ can be written in terms of the Fefferman-Graham expansion of the bulk metric. For example, in d=4+1 dimensions, $\langle T_{ij}\rangle \sim g^{(4)}_{ij}$ \cite{sken}. Then, electromagnetic dual solutions by preservation of the bulk metric, share the same boundary energy.}:
\begin{equation}
g_{\theta\,ij}^{(0)}
=
g_{ij}^{(0)},
\end{equation}
and
\begin{equation}
 \langle T_{ij}\rangle_{\theta}=\langle T_{ij}\rangle.
\end{equation}
However, equality of the metric and stress tensor does not, by itself, imply equality of the complete boundary state. The bulk gauge field carries additional boundary data.

For the electric solution,
\begin{equation}
A_t(r)=\mu-\frac{Q}{r},\qquad\mu=\frac{Q}{r_h}.
\end{equation}
Thus the asymptotic expansion contains both a non-normalizable mode and a normalizable mode, $\mu$ as the boundary source and $-\frac{Q}{r}$ as the normalizable mode. The latter determines the expectation value of the boundary current, up to the normalization appropriate to the holographic dictionary. Under the rotation the electric field strength would be
\begin{align}
F'_{tr}=\frac{Q \cos \theta}{r^2},
\end{align}
By integrating
\begin{align}
A_t^{\theta}=\mu_\theta-\frac{Q \cos \theta}{r}
\end{align}

The gauge choice $A_t(r_h)=0$ results in,
\begin{align}
A_t^\theta=Q \cos \theta \Big(\frac{1}{r_h}-\frac{1}{r}\Big)
\end{align}
where $\mu_\theta=\frac{Q \cos \theta}{r_h}$. For a planar black brane, a constant magnetic field can be written as
\begin{align}
F'_{xy}=-Q e^{-\gamma}\sin \theta
\end{align}
and the potential would be
\begin{align}
A_y^\theta=-Q e^{-\gamma}\sin \theta x
\end{align}

Then, the complete potential is written as
\begin{align}
\mathbf A^\theta=Q \cos \theta \Big(\frac{1}{r_h}-\frac{1}{r}\Big)dt-Q e^{-\gamma}\sin \theta x dy.
\end{align}

At the boundary where $r\rightarrow \infty$
\begin{align}
&A_t^{(0)}(\theta)=\frac{Q \cos \theta}{r_h},\\
&A_y^{(0)}(\theta)=-Q e^{-\gamma}\sin \theta x
\end{align}
the bulk duality rotation generates new asymptotic electromagnetic data, including a magnetic boundary field, while the gravitational boundary metric and stress tensor remain unchanged. Since the original configuration is purely electric, a generic duality rotation produces nonzero magnetic data. Therefore one should distinguish carefully between
\begin{equation}
g_{\theta\,\mu\nu}=g_{\mu\nu},
\qquad
\langle T_{\mu\nu}\rangle_{\theta}=\langle T_{\mu\nu}\rangle,
\end{equation}
and the stronger statement
\begin{equation}
|\Psi_{\theta}\rangle=|\Psi\rangle.
\end{equation}
The latter does not follow from the former alone.

This distinction is particularly relevant for generalized holographic complexity. A geometric functional such as $\mathcal{C}_{Any}(C^2)$ is insensitive to the electromagnetic duality orbit, whereas a matter-sensitive functional such as $\mathcal{C}_{Any}(F^2)$ may distinguish different electromagnetic configurations with identical gravitational geometry.

 \section{Conclusion}\label{con}
In this work, we have investigated the sensitivity of generalized holographic complexity to electromagnetic duality in Einstein-ModMax theory. The electromagnetic duality of ModMax provides a particularly useful setting for this question because configurations related by a continuous duality rotation can have the same gravitational geometry and energy-momentum tensor while differing in their electromagnetic field content. Starting from a purely electric planar AdS black brane, we constructed its duality orbit, which includes mixed electric-magnetic and purely magnetic configurations. Importantly, the metric remains unchanged along this orbit, indicating the gravitational sector's invariance under the duality transformation.

We then studied this duality orbit within the complexity=anything framework. The present analysis shows that the response of holographic complexity to electromagnetic duality depends crucially on the choice of complexity functional. Functionals constructed entirely from curvature invariants, such as the Weyl tensor squared, are insensitive to the duality rotation because they depend only on the unchanged gravitational geometry. In contrast, a matter-sensitive functional involving the electromagnetic invariant $F_{\mu\nu}F^{\mu\nu}$ changes along the duality orbit. Thus, two bulk configurations that are gravitationally indistinguishable can nevertheless be assigned different values of generalized holographic complexity.

To quantify this dependence, we introduced the orbit complexity, which measures the change of a chosen complexity functional under the electromagnetic duality. The orbit complexity vanishes identically for duality-invariant functionals, whereas it is nonzero for functionals that depend on duality-variant electromagnetic invariants. We also introduced the corresponding duality sensitivity, which directly measures how strongly a given generalized complexity functional probes the electromagnetic sector.

Our numerical analysis additionally illustrates this distinction. For the $F_{\mu\nu}F^{\mu\nu}$ functional, both the complexity and its growth rate depend on the duality angle $\theta$, interpolating between the purely electric, mixed, and purely magnetic configurations. Although the qualitative structure of the growth rate can remain similar along the orbit for particular parameter choices, the effective potential demonstrates that its structure, including the number of relevant extrema, can depend on the duality angle. This gives a concrete example in which the choice of matter-sensitive complexity functional affects not only the value of the observable but also its dynamical behavior.

From the holographic perspective, these results illustrate an important aspect of the complexity=anything proposal. The freedom to choose the functional $F_1$ is not simply a freedom in the geometric representation of complexity; it determines which sector of bulk information the observable is sensitive to. In the present example, curvature-based functionals probe information that is invariant under electromagnetic duality, whereas matter-sensitive functionals retain sensitivity to electromagnetic configurations that are degenerate with respect to the gravitational geometry and stress tensor. In this sense, electromagnetic duality provides a useful criterion for organizing generalized complexity functionals according to the duality they preserve.

The boundary interpretation requires some care. Since the bulk metric is unchanged along the duality orbit, the boundary metric source and the holographic stress tensor are likewise unchanged. However, the asymptotic gauge-field data generally rotate under the electromagnetic duality transformation. Consequently, equality of the boundary metric and stress tensor does not by itself imply equality of the complete boundary state. The nontrivial orbit dependence of a matter-sensitive complexity functional therefore indicates sensitivity to boundary duality beyond what the stress tensor encodes, provided the corresponding duality transformation acts nontrivially on the boundary electromagnetic data.

Our results suggest several directions for further investigation. First, it would be interesting to determine whether electromagnetic duality can impose more general constraints on the otherwise large functional freedom of the complexity=anything proposal. Second, one may study analogous orbit complexities in other duality-invariant theories and in higher-dimensional or non-Abelian systems. Finally, understanding the precise boundary interpretation of matter-sensitive complexity, including its relation to the choice of cost function and to the electromagnetic duality action on the boundary Hilbert space, may provide a more systematic characterization of the duality encoded in generalized holographic complexity.




\begin{thebibliography}{}


\bibitem{Bandos:2020jsw}
I.~Bandos, K.~Lechner, D.~Sorokin and P.~K.~Townsend,
``A non-linear duality-invariant conformal extension of Maxwell's equations,''
Phys. Rev. D \textbf{102}, 121703 (2020)
doi:10.1103/PhysRevD.102.121703
[arXiv:2007.09092 [hep-th]].

\bibitem{ModMaxII} B. P. Kosyakov, "Nonlinear electrodynamics with the maximum allowable symmetries," Phys. Lett. B 810 (2020) 135840, arXiv:2007.13878v3 [hep-th].
\bibitem{Bandos2021a}I.~Bandos, K.~Lechner, D.~Sorokin and P.~K.~Townsend, "On p-form gauge theories and their conformal limits,'' JHEP 03 (2021) 022, arXiv:2012.09286.

\bibitem{Bandos2021b}I.~Bandos, K.~Lechner, D.~Sorokin and P.~K.~Townsend, "ModMax meets Susy,'' JHEP 10 (2021) 031, arXiv:2106.07547.
\bibitem{modsol}E. Ayón-Beato, D. Flores-Alfonso and M. Hassaine, “Nonlinearly charging the conformally dressed black holes preserving duality and conformal invariance,” Phys. Rev. D 110, 064027 (2024), arXiv:2404.08753v2 [hep-th].
\bibitem{modsol1}H. M. Siahaan, “Weakly magnetized black holes in Einstein-ModMax theory,” Phys.Lett.B 865 (2025) 139479, arXiv:2409.13967v2 [gr-qc].
\bibitem{RathiRoychowdhury2023}H.~Rathi and D.~Roychowdhury, "AdS$_2$ holography and ModMax,'' JHEP 07 (2023) 026, arXiv:2303.14379v2 [hep-th].
\bibitem{Barrientos2025}J.~Barrientos, N.~C\'aceres, F.~Diaz and U.~Hernandez-Vera, "ModMax electrodynamics and holographic magnetotransport,'' Phys.\ Rev.\ D 112 (2025) 086018, arXiv:2506.02884v2 [hep-th].


\bibitem{Kruglov}S. I. Kruglov,"Thermodynamics of Magnetic Black Holes with Nonlinear Electrodynamics in Extended Phase Space,"  Universe, 10(7), 295.
\bibitem{KRUGLOV2015299}S.~I.~Kruglov, "A model of nonlinear electrodynamics," Annals of Physics, \textbf{353}  (2015), arXiv:1410.0351v4 [physics.gen-ph].
 
 
 
\bibitem{Kar:2024zbo}A.~Kar, "Aspects of a novel nonlinear electrodynamics in flat spacetime and in a gravity-coupled scenario,'' Eur.Phys.J.C 84 (2024) 12, 1246, arXiv:2406.10577 [gr-qc].
 

\bibitem{cv} L. Susskind and D. Stanford, "Complexity and shock wave geometry," Phys. Rev. D 90, 126007 (2014), arXiv:1406.2678[hep-th].
\bibitem{ca} A. R. Brown, D. A. Roberts, L. Susskind, B. Swingle, and Y. Zhao, "Complexity Equals Action," Phys. Rev. Lett. 116, 191301 (2016), 	arXiv:1509.07876 [hep-th].
\bibitem{cv2}J. C., W. Fischler and P. H. Nguyen, ''Noether charge, black hole volume, and complexity," JHEP 1703 (2017) 119, arXiv:1610.02038[hep-th].

\bibitem{TaoWangYang2017}J.~Tao, P.~Wang and H.~Yang, "Testing holographic conjectures of complexity with Born-Infeld black holes,''
Eur.\ Phys.\ J.\ C 77 (2017) 817, arXiv:1703.06297v1 [hep-th].

\bibitem{Meng2019}K.~Meng, "Holographic complexity of Born--Infeld black holes,'' Eur.\ Phys.\ J.\ C 79 (2019) 984, arXiv:1810.02208v4 [hep-th].

\bibitem{PanJing2019} Q.~Pan and J.~Jing, "Holographic subregion complexity in Einstein-Born-Infeld theory,'' Eur.\ Phys.\ J.\ C 79 (2019) 582, arXiv:1808.10169v2 [hep-th].

\bibitem{ShiPanJing2020} Y.~Shi, Q.~Pan and J.~Jing, "Holographic subregion complexity in metal/superconductor phase transition
with Born-Infeld electrodynamics,'' Eur.\ Phys.\ J.\ C 80 (2020) 1100, arXiv:2004.08303.

\bibitem{LingLiuWu2021} Y.~Ling, P.~Liu and J.-P.~Wu, "Complexity for holographic superconductors with the nonlinear
electrodynamics,'' Nucl.\ Phys.\ B 971 (2021) 115615.
\bibitem{general}A. Belin, R. C. Myers, S.-M. Ruan, G. Sárosi, and A. J. Speranza, "Does Complexity Equal Anything?," 
Phys.Rev.Lett. 128 (2022) 8, 081602, arXiv: 2111.02429 [hep-th]


\bibitem{Shepherd:2015dse} 
 B.~L.~Shepherd and E.~Winstanley, "Dyons and dyonic black holes in ${\mathfrak {su}}(N)$ Einstein-Yang-Mills theory in anti-de Sitter spacetime,''
 Phys.\ Rev.\ D {\bf 93}, no. 6, 064064 (2016), arXiv:1512.03010 [gr-qc].
\bibitem{shahbazi}M. Shahbazi and M. Sadeghi, "Callan-Symanzik-like equation in information theory," Eur. Phys. J. Plus 141, 530 (2026), arXiv:2508.13330v3 [quant-ph].
\bibitem{myers}M.-T. Wang, H.-Y. Jiang and Y.-X. Liu, ”Generalized Volume-Complexity for
RN-AdS Black Hole,” JHEP 07 (2023) 178, arXiv:2304.05751v3 [hep-th].
\bibitem{sken}S. de Haro, S. N. Solodukhin and K. Skenderis, "Holographic reconstruction of space-time and renormalization in the AdS / CFT correspondence," Commun.Math.Phys. 217 (2001) 595-622, arXiv: hep-th/0002230.

\end{thebibliography}
\end{document}